# Heavy Fermions in Strongly Correlated Electron System CeAl$_3$

N. E. Sluchanko[1*], V. V. Glushkov[1,2], S. V. Demishev[1,2], N. A. Samarin[1], G. S. Burhanov[3], O. D. Chistiakov[3], D. N. Sluchanko[1].

[1] A. M. Prokhorov General Physics Institute of RAS, 38, Vavilov str., Moscow, 119991, Russia

[2] Moscow Institute of Physics and Technology, 9, Institutskii per., Dolgoprudny, Moscow region, 141700, Russia

[3] Baikov Institute of Metallurgy and Materials Technology of RAS, 49, Leninskii pr., Moscow, 119991, Russia

*-E-mail: nes@lt.gpi.ru

High precision measurements of the Hall effect have been carried out for archetypal heavy fermion compound - CeAl$_3$ in a wide range of temperatures 1.8-300K. For the first time a complex activated behavior of the Hall coefficient in CeAl$_3$ with activation energies $E_{a1}/k_B \approx 125$K and $E_{a2}/k_B \approx 3.3$K has been observed in the temperature intervals 50÷300K and 10÷35K respectively. At temperatures below the maximum of the Hall effect $T<T_{max} \approx 10$K an asymptotic dependence $R_H(T) \sim \exp(-E_{a3}/k_B T)$ was found in CeAl$_3$ with the value $E_{a3}/k_B \approx 0.38$K estimated from the experimental data. The temperature evolution of microscopic parameters (effective mass and localization radius) evaluated for the many-body states (heavy fermions) is discussed in terms of an electron-polaron states' formation in vicinity of Ce-sites in the CeAl$_3$ matrix.

An interest in the anomalous properties of *CeAl$_3$* dates back to 1970 when Buschow et al. [1] found a marked rise in its resistivity at temperatures below $T \approx 250$K. Since the first report of its unusual low temperature specific heat [2] this intermetallic compound is considered to be the most famous and archetypal example of heavy fermion metal. However, in spite of numerous experimental results and scientific publications on this subject, there is no commonly agreed stand of view concerning the origin of many-body states (heavy fermions) in cerium intermetallides up to now.

According to the approach, which is usually applied for cerium intermetallides, the formation of heavy fermions in vicinity of Ce *4f*-centers is explained in terms of the Kondo compensation of Ce$^{3+}$ localized magnetic moments (LMM). As the LMM of rare-earth ion exist in every unit cell of the crystals, such compounds are commonly classified as Kondo-lattices [3]. However, an 'exhaustion problem' noted by some authors' (see, e.g., [4]) arises in case of these dense Kondo systems, where the concentration of charge carriers in conduction band is not sufficient to provide a full LMM compensation when forming the Kondo singlet ground state.

Recent studies of the charge transport carried out for so-called magnetic Kondo-lattice $CeAl_2$ have resulted to another explanation of the nature of heavy fermions in cerium intermetallides [5]. Based on the experimental data the authors of [5] concluded in favor of exciton-polaron complexes forming near the Ce-sites in the $CeAl_2$ matrix and having a typical size $a^*$~10Å. However, the Laves phase compound $CeAl_2$ studied in [5] should be treated as specific one due to strong electron-phonon interaction influencing on its properties [6]. From this point of view, a detailed investigation of charge transport in the archetypal heavy fermion compound $CeAl_3$ seems to be very important to obtain the exhaustive information on charge carriers' characteristics and to shed a light on the nature of many-body states in this model system with fast fluctuations of electron density.

To carry out these high precision measurements of Hall effect, an original scheme using the step-by-step fixation of the sample rotated in steady magnetic field has been applied with a rotation axis perpendicular to the magnetic field [5]. High quality polycrystalline samples of $CeAl_3$ have been used in these investigations performed for a wide range of temperatures 1.8÷300K in magnetic field of permanent magnet ($H_0$~1÷4 kOe). The resistivity of $CeAl_3$ samples has been measured in this study by the standard dc four-point technique.

A family of the angular dependencies of Hall resistance $\rho_H(\varphi)$ obtained from the step-by-step rotation of the $CeAl_3$ sample ($\Delta\varphi\approx3\div8°$) in homogeneous magnetic field ($H_0$=3.7 kOe) is shown in the Fig.1. A cosine behavior of $\rho_H(\varphi)\sim\cos\varphi$, which can be clearly seen from the data of Fig.1, corresponds to the modulation of the normal component of magnetic field $H_\perp=H_0\cos\varphi$ ($H_\perp \| n$, $n$ – normal to the sample surface). The amplitude of the Hall resistivity $\rho_H(T,H_0)$ was used to plot the temperature dependence of Hall coefficient $R_H(T,H_0$=3.7 kOe$)$ in $CeAl_3$ (curve 1 in the Fig.2). It is important to note that according to [7,8] $R_H(T)$ in $CeAl_3$ stays positive in the whole range of 1.8÷300K and is characterized by a wide maximum of prominent amplitude at $T_{max}\approx10K$ (Fig.2). The temperature dependency of resistivity $\rho(T)$ in $CeAl_3$ measured during the experiment is presented in the inset in Fig. 2. From the data the ratio $R_H(T)/\rho(T)=\mu_H(T)$ corresponding to the Hall mobility in case of one-type charge carrier system has been obtained (curve 2 Fig. 2). It should be point out here that the high precision of Hall effect measurements in $CeAl_3$ achieved in these experiments allowed us for the first time to analyze in details the temperature dependencies for $R_H(T)$ and $\mu_H(T)$ parameters and to compare between the obtained data and other characteristics of $CeAl_3$.

Replotting of the $R_H(T,H_0)$ curve in reciprocal semi-logarithmic graph (Fig. 3) allowed us to distinguish between three characteristic temperature ranges of charge transport in $CeAl_3$. Lowering the temperature in the ranges 50÷300K (I) and 10÷35K (II) (Fig. 3a) reveals an anomalous activation type increase of $R_H(T)$ which corresponds to the dependencies

$$R_H(T) \sim \exp(E_{a1,2}/k_BT) \qquad (1)$$

with activation energies $E_{a1}/k_B \approx 125\pm10K$ and $E_{a2}/k_B \approx 3.3\pm0.2K$, respectively. Such an activation type behavior of the Hall coefficient $R_H(T)$ which is very unusual for metallic system has no explanation in framework of the models of Kondo-lattice and skew-scattering [9-10]. Indeed, in these models the spin-flip resonant scattering of conduction electrons on LMM of rare-earth ions is considered as the dominant factor determining charge transport. Within the approach [9-10] both the anomalous positive Hall effect and the non-monotonous behavior of resistivity in heavy fermion compounds (see Fig. 2) should be exclusively ascribed to the specific character of the scattering effects.

In the other approach applied earlier to $CeAl_2$ [5] the $E_{a1,2}$ parameters in (1) can be treated as bound energies of many-body states which form in vicinity of Ce *4f*-centers in the matrix of $CeAl_3$. In this approximation two values of $E_a$ reflect two various situations: (i) when excited levels of the $^2F_{5/2}$-state of cerium are considerably populated (range I) and (ii) when the electronic density fluctuations are induced by transitions into conduction band from the ground $^2F_{5/2}$-state doublet (range II). This suggestion is supported by the scheme of CF splitting of cerium $^2F_{5/2}$-state obtained from inelastic neutron scattering experiments [11] (see inset in Fig.3). Thus, at intermediate temperatures 50÷300K the formation of many-body states in $CeAl_3$ matrix is strongly influenced by inelastic processes. In this temperature range the fast spin/charge fluctuations leading to the charge carriers' polarization occur between the populated *4f*-doublets and conduction band states. According to [11] the excited doublets of Ce $^2F_{5/2}$-state turn out to be considerably broadened and form a band of excited states of the width $\Gamma$ ($\Gamma/2 \approx E_{a1,2}/k_B \approx 125K$) which is the same order as the splitting value of $\Delta_2 \approx 170K$ (inset in Fig. 3).

The value of the width of quasi-elastic peak $\Gamma_0(T)/2 \approx 5K$ obtained from the neutron scattering experiments in $CeAl_3$ [11] was used within simple relation

$$\Gamma_0(T)/2 = \hbar/\tau_{eff}(T) \qquad (2)$$

to evaluate the relaxation time $\tau_{eff}(T)$ and, further on, to estimate the value of effective mass $m^*$ of heavy fermions

$$m^*(T) = e\tau_{eff}(T)/\mu_H(T) \qquad (3).$$

The calculated values of $m^*_1(77K) \approx 90m_0$ and $m^*_2(20K) \approx 45m_0$ are consistent very well to the effective masses of spin-polaron and exciton-polaron states estimated for other strongly correlated electron systems FeSi ($m^* \approx 20 \div 90 m_0$ [12]), $SmB_6$ ($m^* \approx 20 \div 40 m_0$ [13]) and $CeAl_2$ ($m^*_{1,2} \approx 55 \div 90 m_0$ [5]). Moreover, we also used the parameters $E_{a1,2}$ and $m^*_{1,2}$ to estimate the localization radius of the many-body states

$$a^*_{p1,2} = \hbar / (2E_{a1,2} m^*_{1,2})^{1/2} \qquad (5).$$

For heavy fermions in CeAl$_3$ matrix this estimation gives the values of 1.8Å and 16.4Å in ranges I and II, respectively. It is necessary to note here a small value of $a^*_p$ in range I (spatial size of nanoregions $a^*_{p1} \approx 1.8$Å $< a$, where $a$ is the lattice constant) corresponding to a "deep potential well" on the Ce$^{3+}$-sites ($E_{a1}/k_B \approx 125 \pm 10$K, range I in Fig.3a). Such a small characteristic size (~1.8Å) of many-body states in CeAl$_3$ cannot be explained within the Kondo lattice model. On the contrary, the significant 'delocalization' of the many-body states ($a^*_{p2} \approx 16.4$Å) observed at low temperatures is accompanied by a dramatic decrease of the bound energy ($E_{a2}/k_B \approx 3.3 \pm 0.2$K, range II in Fig.3a). In this respect it is naturally to assume that this essential increase of the localization radius induces a further transition to coherent state, which is displayed in the abrupt decrease of resistivity and Hall coefficient in rare-earth intermetallides with strong electron correlations. Additionally, at temperatures below 30K the temperature dependence of Hall mobility in CeAl$_3$ can be well fitted by an exponent $\mu_H \sim T^\alpha$ with $\alpha = 0.35 \pm 0.04$ (curve 2 in Fig.2). So the observed discrepancy between the experimental data and the Curie-Weiss dependence of Hall mobility $\mu_H^{-1}(T) \sim (T+\theta) \sim \chi^{-1}(T)$ predicted within skew-scattering model [9-10] can be treated as another argument in favor of proposed exciton-polaron approach [5].

Another novel experimental result of this study can be observed at temperatures below the maximum of Hall effect in CeAl$_3$ (range III, $T < T_{max} \approx 10$K) where the temperature variation of $R_H$ follows to the dependence

$$R_H(T) \sim \exp(-E_{a3}/k_B T) \qquad (4)$$

with the estimated value of $E_{a3}/k_B \approx 0.38 \pm 0.02$K (Fig.3b). It is interesting to note here that such a behavior of Hall coefficient was earlier predicted for systems with topologically nontrivial spin background [14-15]. As shown in [14-15] Hall effect in the systems is drastically influenced by the effective internal magnetic field $H_{int} = \langle h_z \rangle \sim (1/k_B T) \exp(-E_a/k_B T)$ arising from Berry phase contribution. This appearance of $H_{int}$ in nano-size regions around the Ce-sites in CeAl$_3$ seems to be also supported by the estimations of induced magnetic moment on the Ce-sites $\mu_{eff} \approx 0.05 \mu_B$ obtained from the data of μSR and NMR studies [16-17].

To summarize, the study of Hall effect performed for archetypal heavy fermion compound CeAl$_3$ allowed us to obtain the temperature evolution of microscopic parameters (effective masses and localization radii) of the many-body states. It was shown from the experimental data that the short-range electron-polaron states (see, e.g., [18-19]) formed in the regime of fast spin/charge fluctuations in the vicinity of cerium ions in studied intermetallides should be allowed for an adequate interpretation of the low temperature transport anomalies.

This work was financially supported by projects RFBR 04-02-16721 and INTAS 03-51-3036 as well as by the Program "Strongly Correlated Electrons in Semiconductors, Metals, Superconductors, and Magnetic Materials" of RAS and by the Program "Development of High School Scientific Potential" of MES RF. One of us (V.V.G.) acknowledges personal support from Russian Science Support Foundation.

**Figure Captions.**

**Fig. 1.** Angular dependencies of the Hall resistance $\rho_H(\varphi, T_0)$ of CeAl$_3$ measured in magnetic field $H_0 \approx 3.7$ kOe at different temperatures.

**Fig. 2.** Temperature dependencies of Hall coefficient $R_H$ (curve 1) and Hall mobility $\mu_H(T) = R_H(T)/\rho(T)$ (curve 2) in CeAl$_3$. Inset shows the temperature dependence of resistivity in CeAl$_3$.

**Fig. 3.** Temperature dependence of the Hall coefficient of CeAl$_3$ presented in reciprocal semi-logarithmic graph (a) for the ranges I and II and (b) for the range III (see text). Inset shows the crystal field (CF) splitting of cerium $^2F_{5/2}$-state in CeAl$_3$ [11].

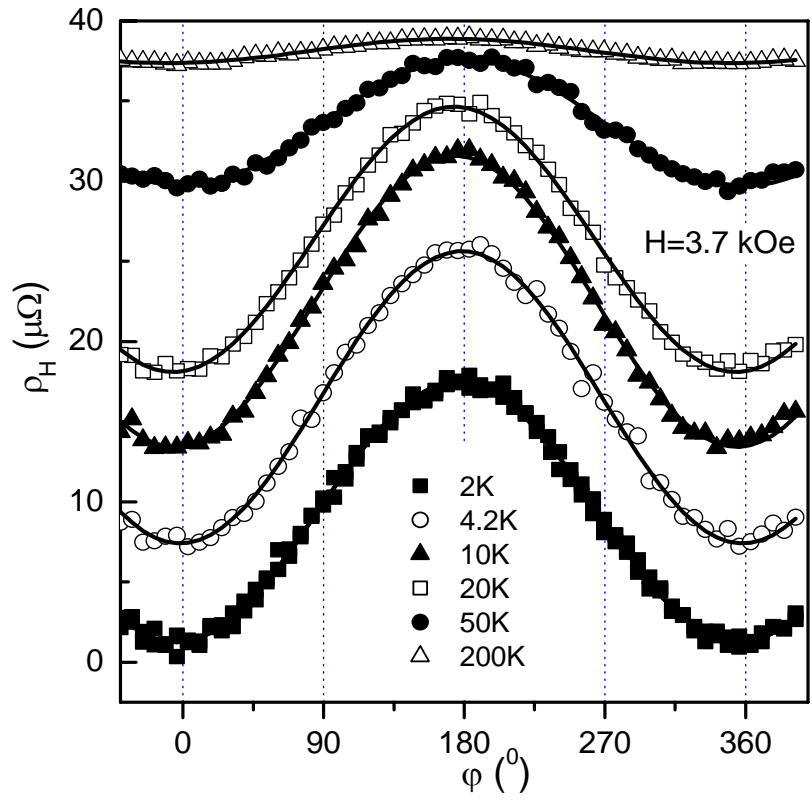

Fig.1

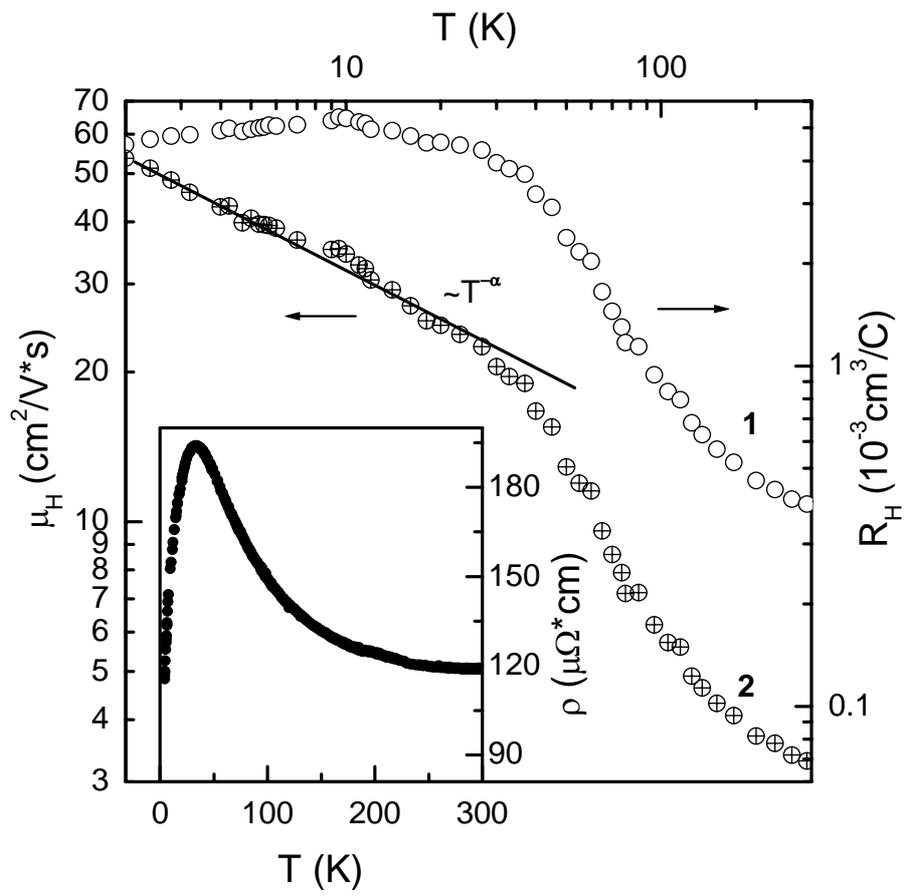

Fig.2

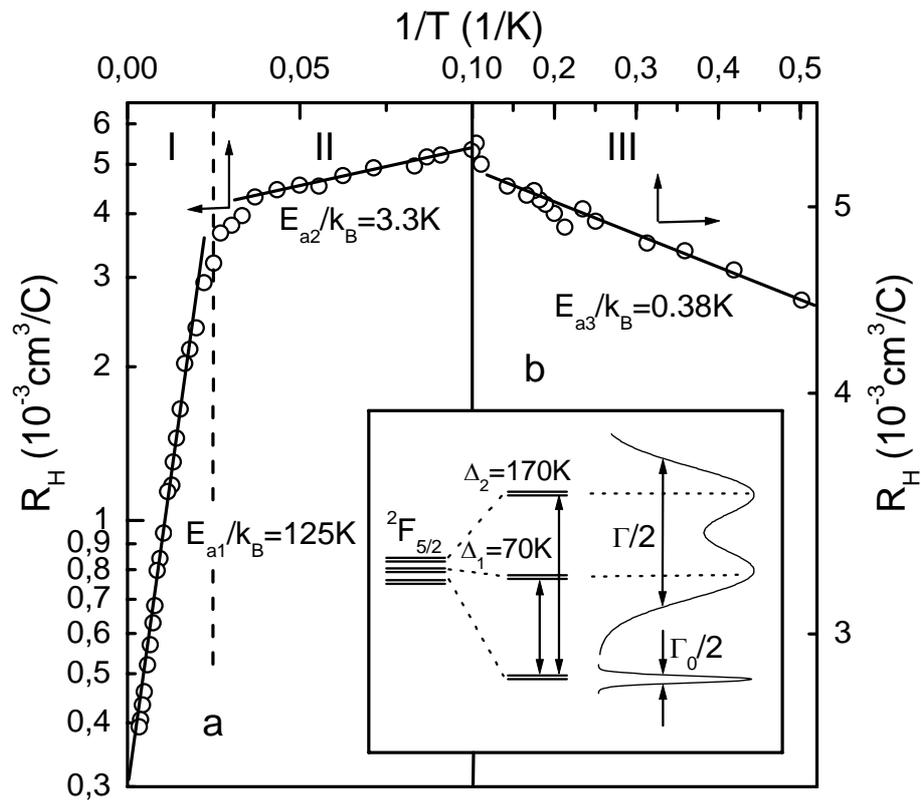

Fig.3